\documentclass[%
reprint,
superscriptaddress,
 amsmath,amssymb,
 aps,
]{revtex4-2}

 \usepackage{amsmath}
 \usepackage{amsfonts}
 \usepackage{amssymb}
\usepackage{graphicx}
\usepackage{fancyhdr}
\usepackage[sc, small]{caption}
\usepackage{lmodern}
\usepackage{color, pstricks}
\usepackage{tcolorbox}
\usepackage{booktabs}  
\usepackage[table,xcdraw]{xcolor}
\usepackage{multirow}
\usepackage{lipsum}
\usepackage{float}
\usepackage{dcolumn}
\usepackage{hyperref}
\hypersetup{
     colorlinks=true,
     linkcolor=black,
     filecolor=magenta,      
     urlcolor=blue,
     citecolor=blue,
     citebordercolor=blue,
     pdfpagemode=FullScreen,
     }

\begin{document}

\title{Self-organized Layered Structures of Nitrogen-Vacancy Centers
with Preferential Orientation in Heteroepitaxial Diamond Films}

\author{Vadim Lebedev}
\email{Corresponding author: vadim.lebedev@iaf.fraunhofer.de}
\affiliation{Fraunhofer Institut für Angewandte Festkörperphysik IAF, Tullastraße 72, 79108 Freiburg, Germany}

\author{Alejandro Martínez-Méndez}
\author{Jesús Moreno-Meseguer}
\affiliation{Departamento de Física—CIOyN, Universidad de Murcia, Murcia E-30071, Spain}

\author{Jan Engels}
\author{Jürgen Weippert}
\author{Jan Kustermann}
\author{Felix Hoffmann}
\affiliation{Fraunhofer Institut für Angewandte Festkörperphysik IAF, Tullastraße 72, 79108 Freiburg, Germany}

\author{Javier Prior}
\affiliation{Departamento de Física—CIOyN, Universidad de Murcia, Murcia E-30071, Spain}

\author{Jan Jeske}
\affiliation{Fraunhofer Institut für Angewandte Festkörperphysik IAF, Tullastraße 72, 79108 Freiburg, Germany}




    


\begin{abstract}

Here, we report the fabrication and characterization of single crystal diamond films on a
two-inch wafer-scale containing three-dimensional layered structures of negatively charged
nitrogen-vacancy color centers (NV$^{-}$). The (100)-oriented layered diamond films were
grown by chemical vapor deposition (CVD) to create spatially separated regions of enhanced
emission associated with NV$^{-}$ centers. Such localized emission enhancement as well as 2D
and 3D arrays of spins could enable close-packed architectures for emerging quantum
devices, such as data processors, memory cells, and magnetic sensors. A cyclic sequence of
CVD growth and oxygen-plasma etching is employed, resulting in the in-situ formation of
layered inverted pyramidal structures with flat $\{$111$\}$ sidewalls. These pits are subsequently
overgrown by nitrogen-doped diamond to achieve NV$^{-}$-rich facets with a uniform N–V bond
arrangement and improved nitrogen incorporation efficiency.
\end{abstract}

\maketitle


\section{Introduction}
Nitrogen-vacancy (NV) centers are point defects in the diamond (Fd3m) crystal lattice
(cC). In the negatively charged state (NV\-), they possess an electron spin triplet and can
therefore assume three possible spin states. These spin states are extremely stable and can
be initialized, controlled, and read out at room temperature \cite{ref01}. Accordingly, NV centres
have attracted considerable interest for both quantum sensing, as atomically small or
sensitive ensemble sensing elements, and for quantum computing, where they serve as solid-state qubits \cite{ref02}. Growth along the cC[111] direction is particularly attractive because NV
centers are preferentially incorporated along this growth direction, leading to enhanced
sensitivity of the NV\- centers to magnetic fields aligned with the same crystallographic axis
 \cite{ref03}. This is also crucial for qubit addressability in quantum computing and data storage
devices.

High-quality cC(111) can be grown on iridium/ yttria stabilized zirconia (YSZ)/ silicon (111)
templates with an NV-pair alignment close to 100\% along the growth direction \cite{ref04, ref05, ref06}. In particular, bias-enhanced nucleation (BEN) and CVD growth of 2-inch cC(111) wafers have
already been reported by our joint team \cite{ref07}. Our highest-quality NV-doped heteroepitaxial
epilayers exhibit a T$_2$ coherence time of 26 µs at a nitrogen concentration of 1 ppm and a
full width at half maximum (FWHM) of the (111) reflection of FWHM(111) $<$ 0.5° \cite{ref08}. The
total nitrogen concentration was determined to be 7.3 ppm, which, together with
photoluminescence (PL) analysis, indicates an NV incorporation efficiency of
approximately 0.1\%. Nevertheless, the major obstacle to the production of cC(111)
heteroepitaxial material remains the extremely low growth rate of 100–300 nm/h, which
substantially increases the material cost.

In contrast, the CVD growth and post-growth processing of (01) heteroepitaxial films
nucleated and grown on Ir(100) templates with diameters up to 2 inches have been widely reported in the literature \cite{ref09}. To date, (100)-oriented CVD diamond growth can be considered a mature technology, resulting in FWHM(002) $\ll$ 0.1° for thick epilayers. High spin coherence times of T$_2$ $>$ 1 ms and T$_2$$^*$ $>$ 100 µs have been reported for a nitrogen
concentration in cC films of 1 ppm at room temperature. However, due to the random
orientation of NV pairs (four magnetically non-equivalent orientations), such epilayers
exhibit only limited contrast for each orientation because only one quarter of the NV centres
contribute to the signal for a given orientation, while the remaining three quarters contribute
background fluorescence. The fabrication of aligned NVs along a preferential crystal
orientation can lead to a strong enhancement of the contrast and thus sensitivity for one
orientation as well as the ability to measure this orientation without the need to apply a bias
magnetic field to make the four orientations individually addressable.

An alternative fabrication method involving selective growth of cC(111) material using
profiled cC(100) samples was demonstrated by Schreyvogel et al. in 2021 \cite{ref10}. They showed
that cC(111) can be selectively grown on $\{$111$\}$ facets of pyramidal pits formed by catalytic
Ni-based etching of the (100) surface. These $\{$111$\}$ facets were subsequently overgrown with
phosphorus-doped diamond film to obtain highly n-type material on the sidewalls.

In this work, we demonstrate that, using a recently developed cyclic CVD process (cCVD)
\cite{ref11, ref12, ref13} consisting of alternating CVD deposition and oxygen-plasma etching steps,
reproducible 3D light-emitting structures can be created in bulk cC(100) material. The
samples were grown on 2-inch epitaxial iridium/YSZ/silicon (100) wafers \cite{ref14}. As-grown
300 µm-thick (100) diamond wafers were liberated from the silicon substrate using a KOH
solution and polished on both sides using conventional mechanical polishing. The 2-inch
wafers were diced into 10 × 10 mm² pieces for subsequent characterization, by scanning
electron microscopy (SEM, Jeol JSM-7610F) equipped with cryogenic  cathodoluminescence (CL, Gatan), as well as quantum-optical characterization, such as optically detected magnetic resonance (ODMR) and confocal photoluminescence (cPL, Renishaw InVia) using 532 nm laser excitation. Details of the experimental procedures are provided in previous works \cite{ref08, ref15}.
\vspace{-10pt}

\section{Methods}
\begin{figure}[h]
  \centering
  \includegraphics[width=0.4\textwidth]{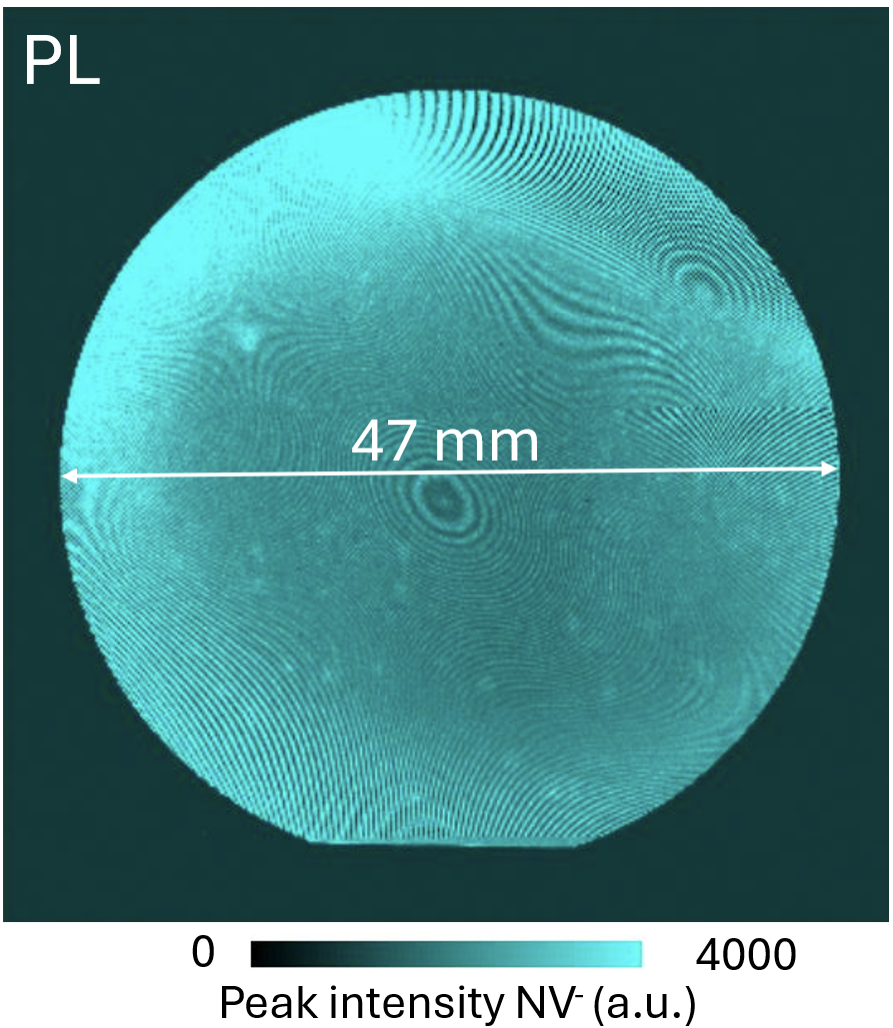}  
  \caption{Photoluminescence (PL) map of the full 47 mm-diameter diamond wafer. Shown is the peak intensity of the zero-phonon-line (ZPL) associated with the negatively charged (NV$^{-}$) state of the NV center, corrected by the baseline from the phonon side bands.}
  \label{Fig:PLmap}
  \vspace{-5pt}

\end{figure}

Fig. \ref{Fig:PLmap} presents a photoluminescence (PL) map of the full-area (001) both-sided polished
diamond wafer acquired at an excitation power of 5 mW and a pixel dwell time of 80 ms.The spatial contrast observed across the wafer originates from interference fringes caused
by multiple internal reflections of the PL emission between the wafer surfaces. Gradients
towards the edges may either originate from actual concentration differences or be an effect
of sample focus.

\begin{figure}[h]
  \centering
  \includegraphics[width=0.4\textwidth]{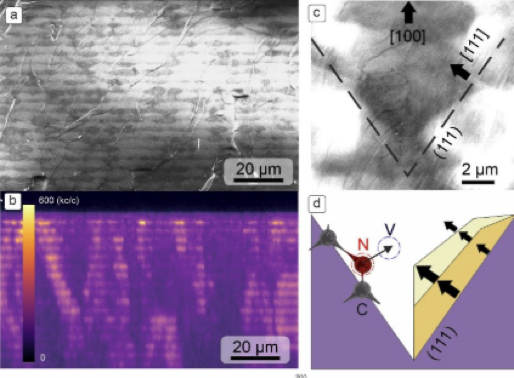}
  \caption{Cross-sectional studies of the sample prepared by a cCVD process: cross-sectional a) SEM and b) cPL images demonstrate variations in a NV incorporation efficiency. c) High-resolution, cross-sectional SEM showing the structure of a $\{$111$\}$-bounded pit, d) schematic illustration of a (111) growth mode along with the dominant orientation of the NV pairs.}
  \label{Fig:Fig2}
  \vspace{-7pt}
\end{figure}

SEM (Fig. \ref{Fig:Fig2}.a) and cPL (Fig. \ref{Fig:Fig2}.b) images show closely matching contrast features recorded
by the secondary-electron detector and in the NV$^{-}$-emission intensity recorded by cPL at the
wavelength of 637 nm (zero-phonon line, ZPL). A high-resolution cross-sectional SEM
image (Fig. \ref{Fig:Fig2}.c) reveals the structure of these emitting spots more precisely. We propose that
the emission originates from $\{$111$\}$-bounded structures formed within the (100) bulk
material. A schematic representation of such a structure, derived from SEM and cPL
observations, is displayed in Fig. \ref{Fig:Fig2}.d.

\vspace{-2pt}

We propose the following mechanism to explain the formation of this pattern during the
cCVD process: i) reactive oxygen-plasma etching forms characteristic $\{$111$\}$-bounded pits,
whose shape is defined by the anisotropy of the etch rates in the $<$111$>$ and $<$111$>$ directions;
ii) subsequent overgrowth of the pits by CVD promotes (111)-oriented growth within the
pits, enhancing nitrogen doping efficiency and promoting the preferential orientation of
[111]-aligned NV pairs.

\begin{figure}[h]
  \centering
  \includegraphics[width=0.4\textwidth]{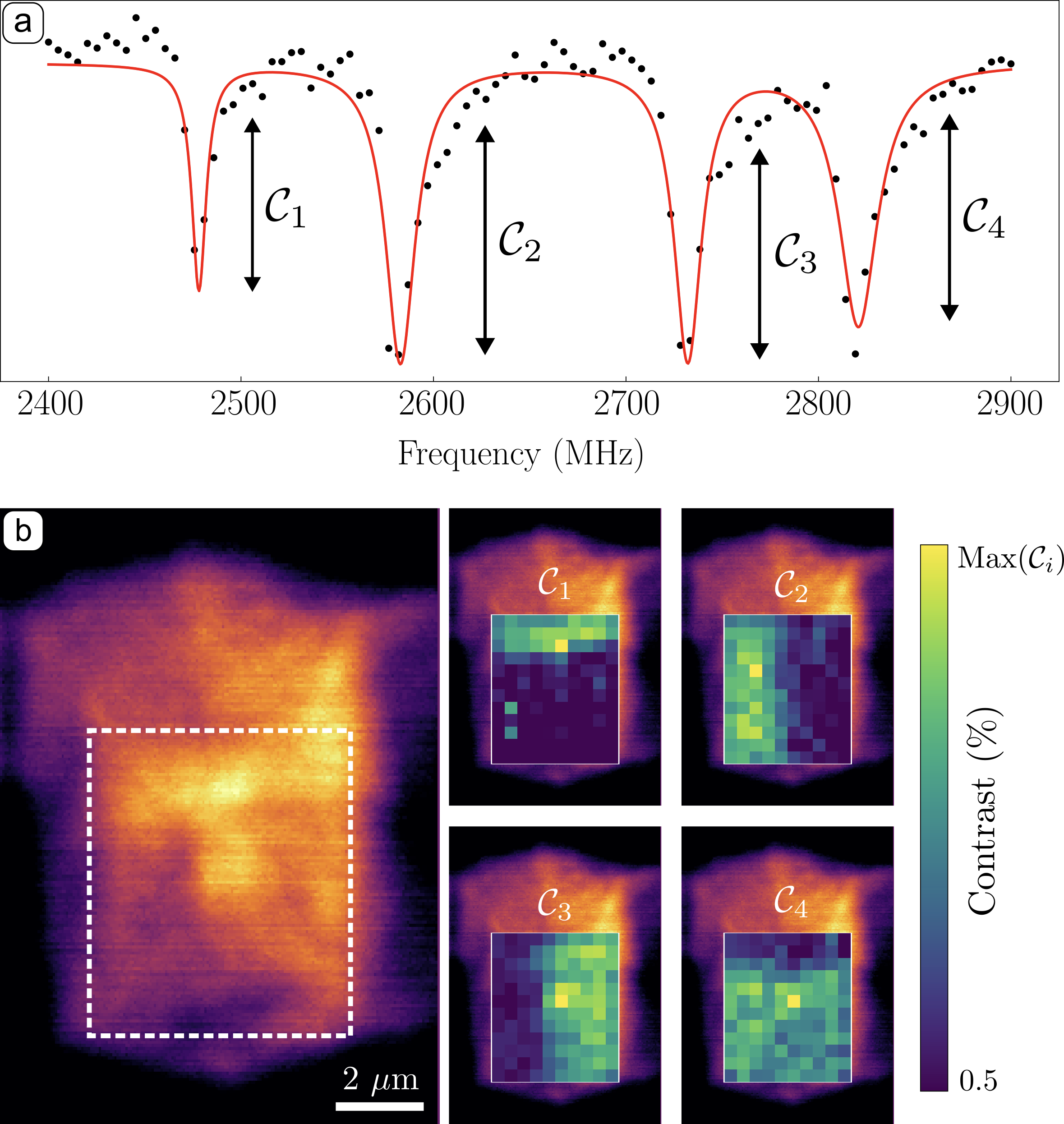}
  \caption{ODMR study of NV$^-$ populated areas. a) An offset magnetic field is added to separate the electronic spin transitions of each of the four crystallographic directions. b) the optical contrast $C_i$ of each dip is obtained from a 10x12 pixel region (zone under white rectangle), revealing four complementary regions of preferential NV alignment.}
  \label{Fig:odmr}
  \vspace{-7pt}
\end{figure}

To confirm the selective (111)-oriented growth within the pits, ODMR mapping was
carried out to identify regions of preferential alignment of NV$^{-}$.
An external magnetic field was applied to separate the spin transitions of the four crystallographic directions, and the optical contrast of each dip was measured (Fig. \ref{Fig:odmr}.a). Four complementary regions of preferential alignment are observed (C1-C4), confirming the presence of $\{$111$\}$-bounded structures.Fig. \ref{Fig:odmr}.b shows that each of the four $\{$111$\}$ sidewalls demonstrates nearly uniform alignment of NV pairs along the [111] growth direction. Such a placement of NV centers has
already been used by our colleagues in focused-ion-beam (FIB) prestructured homoepitaxial diamond \cite{ref16}.

\begin{figure}[h]
  \centering
  \includegraphics[width=0.48\textwidth]{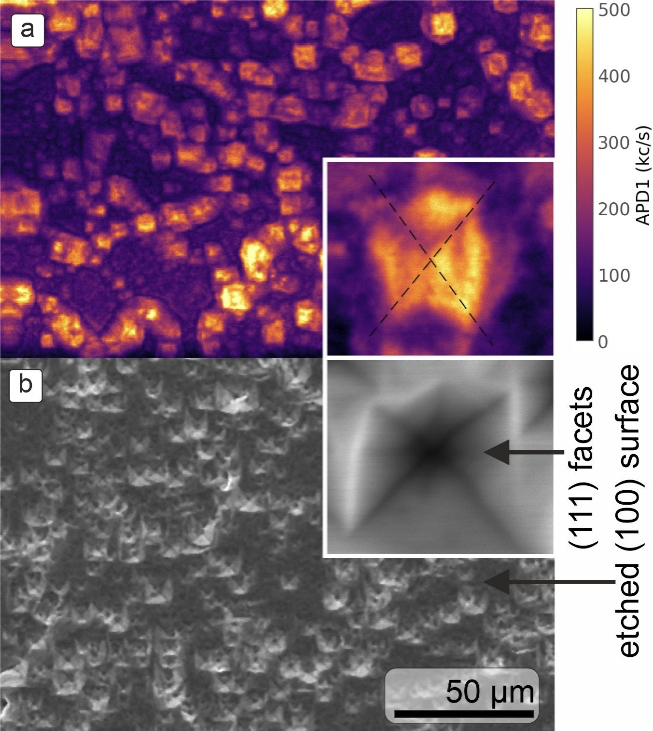}
  \caption{Plan-view studies of the samples prepared by the cCVD process: a) PL map of the overgrown surface (insert: high-resolution PL map); and b) SEM image of the oxygen-plasma-etched (100) surface with the characteristic {111}-bounded pits (inset: high-resolution SEM image).}
  \label{Fig:pyramids}
  \vspace{-5pt}
\end{figure}

Fig. \ref{Fig:pyramids} displays the results of plan-view studies of samples prepared by the cCVD process:
a PL map of the overgrown surface in Fig. \ref{Fig:pyramids}.a and an SEM image of the oxygen-plasma-etched (100) surface with characteristic $\{$111$\}$-bounded pits in Fig. \ref{Fig:pyramids}.b. High-resolution PL and SEM images demonstrate the internal structure of the emitting pit features. These
observations support the model proposed in Fig. \ref{Fig:Fig2}.d: the enhanced NV$^{-}$ emission originates from the inclined sidewalls of the pits overgrown by the diamond epilayer, presumably via
a (111) step-flow growth mode. This interpretation is consistent with the well-established
observation that such a growth mode leads to enhanced doping efficiency and high
uniformity of NV-pair alignment compared with conventional planar (100) CVD growth.

\begin{figure}[h]
  \centering
  \includegraphics[width=0.48\textwidth]{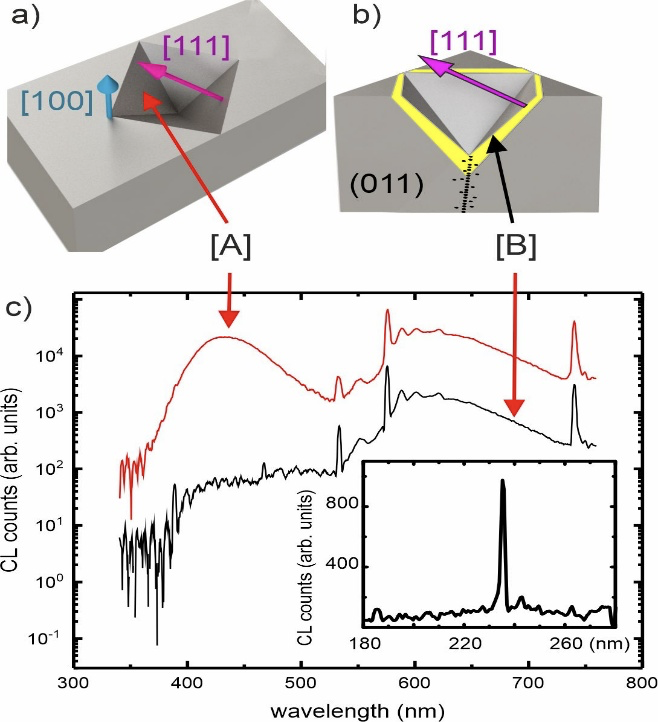}
  \caption{Model of the proposed (111) overgrowth mechanism. (a) Isometric view of a {111}-bounded pit after oxygen-plasma etching. (b) Cross-sectional schematic illustrating the formation of (111)-oriented epilayers on the {111} sidewalls. (c) CL spectra acquired from the etched pit (position A) and from the overgrown surface (position B).}
  \label{Fig:spectrum}
\end{figure}

CL studies performed at 77 K (Fig. \ref{Fig:spectrum}) typically do not allow detection of the NV$^{-}$ related
signal. Here, only the neutral NV$^0$ state at 575 nm can be observed. However, CL clearly
demonstrates the effect of overgrowth on the resulting crystal quality. In Fig. \ref{Fig:spectrum}.c, two spectra are shown, acquired from the oxygen-plasma-etched region (position [A]) directly after the
etching process and on the overgrown high-quality surface (position [B]).

Notably, the CL peak related to free excitons (XTO, see inset in Fig. \ref{Fig:pyramids}.c) is clearly
observable at 77 K, indicating the high quality of the heteroepitaxial diamond. The only
significant difference between the spectra [A] and [B] is the appearance of a broad band
centered around 440 nm (the so-called A-band) in the etched samples. In the literature, this
“blue” emission is associated with a variety of structural defects in diamond \cite{ref17}. In our
case, this band is most likely attributed to vibronic transitions associated with dislocation-
related defects \cite{ref18, ref19}. The most likely origin of this emission is edge-type threading
dislocations with a Burgers vector of b = ½[110]$|$(001). According to our previous studies 
\cite{ref20}, these defects form the densest dislocation network in heteroepitaxial (100) epilayers
and constitute the core of the etched pits, as shown in Fig. \ref{Fig:pyramids}.b.

\begin{figure}[h]
  \centering
  \includegraphics[width=0.45\textwidth]{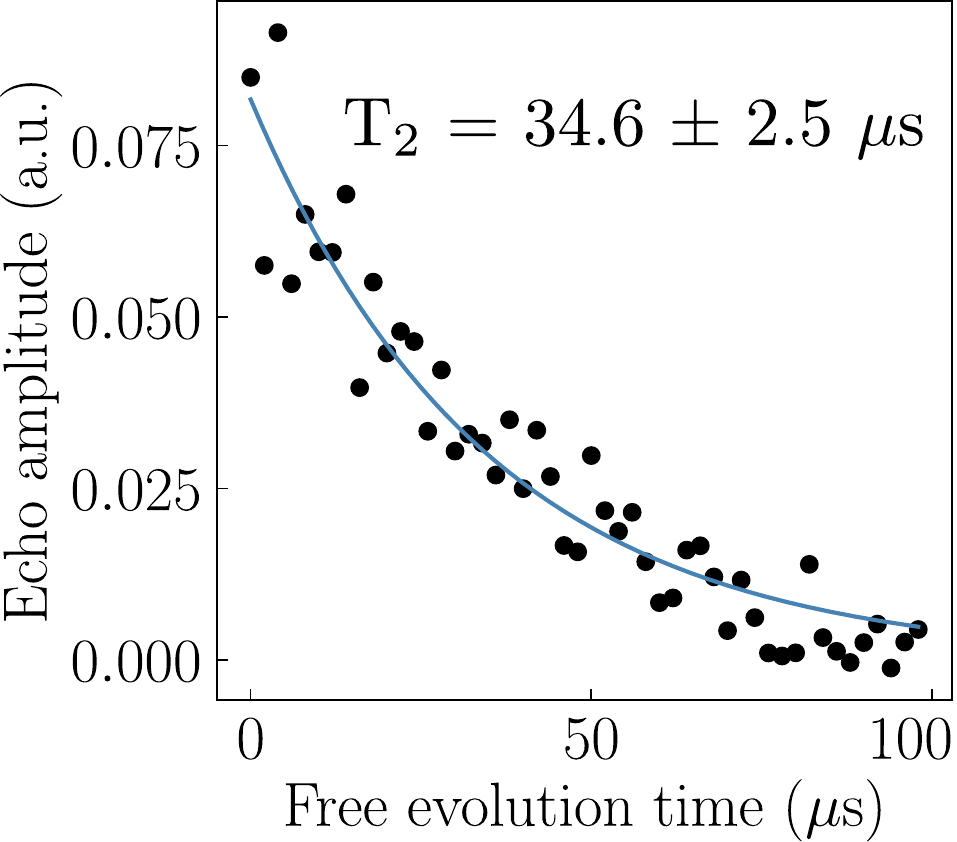}
  \caption{Hahn-Echo measurement of the coherence time T$_2$ = 34.6 $\pm$ 2.5 µs.}
  \label{Fig:echo}
\end{figure}

In addition, a Hahn-Echo sequence - shown in Fig. \ref{Fig:echo} - was carried out to determine the coherence time T$_2$ = 34.6 ± 2.5 µs. In combination with the increased optical contrast from the
preferential alignment, this value provides a strong starting point for these $\{$111$\}$-bounded
structures to be used in different sensing schemes.

\vspace{5pt}

\section{Conclusion}

In summary, a series of growth experiments were carried out to demonstrate that layered
NV$^{-}$ -enriched micro-clusters self-assemble in epitaxial diamond using the cyclic CVD
method. Such clusters form reproducible patterns of enhanced emission spots with the NV$^{-}$
ZPL at 637 nm. The geometry of these 3D patterns can be controlled by varying the
etching/growth cycle parameters. The mechanism responsible for the formation of such NV$^{-}$-rich clusters is the enhanced overgrowth of $\{$111$\}$ facets bounding the etched pits on the
(100) surface. These etch pits are reproducibly formed during oxygen-plasma treatment of
the diamond surface. Most pits contain threading dislocations at their core, which define
their location and properties.

Such two- and three-dimensional addressable arrays of NV$^{-}$ emitters could be employed in emerging quantum technologies for a variety of applications, for example, in vector
magnetometry or wide-field microscopy. The 2D arrays of spins could enable investigations
of out-of-equilibrium phase transitions or time crystals in such arrays \cite{ref21}. In combination
with further processing steps that localize color centers in the x-y plane the given layers
could represent an important step in the creation of arrays of addressable color centers for
quantum computing and communication. Furthermore, the observations show an interesting
mechanism to create preferentially aligned NV centers in large layers of (100)-oriented
diamond. A full characterization of the NV centers’ sensitivity and electronic and nuclear
environment is needed to determine their viability for these purposes.

In many applications both the positioning and the geometrical and emission properties of
the NV centers must be precisely controlled. Accordingly, future work should focus on the
deterministic positioning of these clusters using lithographic techniques, particularly iridium
masks for selective etching and deposition.

\begin{acknowledgments}
J.J. acknowledges funding from the Bundesministerium für Bildung und Forschung (Grant
No. 13N16485). J.P. is supported by the QuantERA II program (Mf-QDS) and QuantERA
III program (AQuSeND) that have received funding from the European Union’s Horizon
2020 research and innovation program under Grant Agreement No. 101017733 and from the
Agencia Estatal de Investigación, with project codes PCI2022-132915, PCI2024-153474
and by QNAVIUM Project SCPP2400C011413XV0 funded by
MICIU/AEI/10.13039/501100011033 and by the European Union
NextGenerationEU/PRTR.
.
\end{acknowledgments}\textbf{}

\bibliographystyle{IEEEtran}
\bibliography{pyramids.bib}

\newpage

\end{document}